\documentclass[floatfix,10pt,superscriptaddress,twocolumn,showpacs,showkeys,aps,pre,final]{revtex4-2}

\usepackage{ifpdf,braket}
\usepackage{amsmath}
\usepackage{amssymb}
\usepackage{graphicx}

\usepackage{lineno}

\usepackage{amsfonts}%
\usepackage{amssymb}%
\usepackage{enumitem}%

\usepackage{textcomp}
\usepackage{ifpdf}

\usepackage{multirow}

\begin{document}

\title{Evidence of the de Almeida-Thouless transition in three-dimensional spin glasses}

\author{L.~H.~Miranda-Filho}
\email[]{lucmiranda@gmail.com}
\affiliation{Institute of Theoretical Physics, Chinese Academy of Sciences, Beijing 100190, China}

\author{Yuliang Jin}
\email[]{yuliangjin@mail.itp.ac.cn}
\affiliation{Institute of Theoretical Physics, Chinese Academy of Sciences, Beijing 100190, China}
\affiliation{School of Physical Sciences, University of Chinese Academy of Sciences, Beijing 100049, China}
\affiliation{Wenzhou Institute, University of Chinese Academy of Sciences, Wenzhou, Zhejiang 325000, China}

\date{\today}

\begin{abstract}

The nature of  spin-glass states in a magnetic field remains a major open problem in statistical physics. The existence of the de Almeida-Thouless (dAT) transition for three-dimensional (3D) spin glasses in a field is still debated.
We introduce a new computational method to define the spin-glass susceptibility, which is robust against the broad tail in  the overlap distribution that undermines conventional analyses. Applying this approach to the Edwards-Anderson spin-glass model in 2D and 3D, and contrasting with the 3D Ising (without disorder) and mean-field spin-glass models, we find a stark difference: the locus of susceptibility maxima bends to the right in the field-temperature plane for the Ising and 2D spin-glass cases, indicating a supercritical crossover line, but bends to the left for the mean-field and 3D spin glasses - a  signature of the dAT line. Finite-size scaling further suggests that the peak susceptibility diverges with system size in 3D spin glasses under a field, while saturating in 2D. These results provide direct numerical evidence for the dAT transition in 3D, supporting the replica symmetry breaking scenario.

\end{abstract}

\maketitle

\section{Introduction}
Extensive theoretical~\cite{bray1980renormalisation, pimentel2002spin, urbani2015gardner, yeo2015critical, singh2017almeida, charbonneau2017nontrivial, charbonneau2019morphology, holler2020one, angelini2022unexpected, angelini2025critical}, numerical~\cite{marinari1998critical, krzakala2001zero, young2004absence, sasaki2007scaling, jorg2008behavior, katzgraber2009study, leuzzi2009ising, baity2014three, baity2014dynamical, dilucca2020spin, munoz2020learning, paga2021spin},
and experimental~\cite{aruga1994experimental, mattsson1995no, petit1999ordering, petit2002ordering, jonsson2005dynamical, tabata2010existence} efforts have been devoted to understanding the spin-glass transition in a finite magnetic field ($h>0$) (see Refs.~\cite{martin2023numerical, altieri2024introduction, dahlberg2025spin} for recent reviews). In the mean-field Sherrington-Kirkpatrick (SK) model~\cite{sherrington1975solvable}, the transition between the spin-glass and paramagnetic phases, occurring at the de Almeida-Thouless (dAT) line in a field-temperature ($h$-$T$) phase diagram, was established theoretically nearly fifty years ago~\cite{de1978stability}. For finite-dimensional systems below the upper critical dimension, however, the situation proves significantly more complex.

Searches for the dAT transition in three dimensions (3D) remain inconclusive, as conventional finite-size scaling (FSS) analyses of the dimensionless correlation length fail to show the intersection point characteristic of a continuous phase transition~\cite{young2004absence, jorg2008behavior, vedula2024evidence, vedula2025erratum, vedula2023study, vedula2025nature}. However, the reliability of this conventional approach has been called into question. Recent work highlights its potential inadequacy, attributing it to large scaling corrections arising from anomalies in  the zero wave-vector mode~\cite{banos2012thermodynamic}, and the masking of typical system behavior in averaged quantities due to strong sample-to-sample fluctuations~\cite{baity2014three, parisi2012numerical}. In contrast, evidence for the dAT line has been shown in higher dimensions, specifically for $d \geq 4$~\cite{banos2012thermodynamic, aguilar2024evidence, aguilar2025evidence}.

In the absence of a magnetic field ($h=0$),
the spin-glass transition has been well-established for $d \geq 2$. In 2D, the transition occurs at zero temperature, $T_{\rm c} = 0$~\cite{morgenstern1979evidence, jain1986monte, cheung1983equilibrium, bhatt1988numerical}, while in $d \geq 3$, $T_{\rm c}$ is finite~\cite{ballesteros2000critical, jorg2008universality, katzgraber2006universality}.
Recently, this scenario has been reinforced by conformal invariance within a scattering field-theoretic framework that implements the replica method~\cite{delfino2026nature}. In 2D, the transition does not occur at a finite $T_{\rm c}$ due to the internal continuous symmetry, which gives rise to a line of renormalization-group fixed points. Above 2D, the continuous symmetry can be broken spontaneously, yielding a transition at a finite $T_{\rm c}$.

\begin{figure*}[!htbp]
  \centering
  \includegraphics[width=\linewidth]{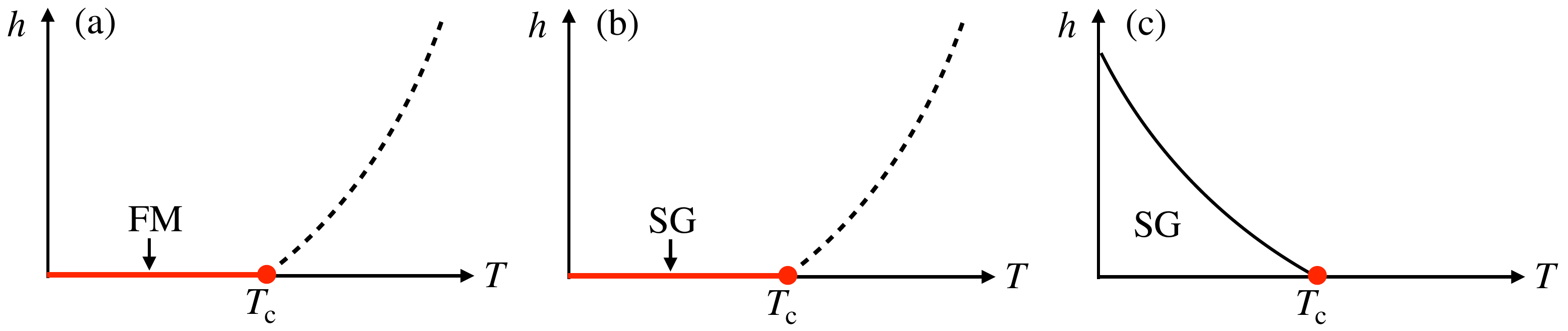}
  \caption{{\bf Schematic $h$-$T$ phase diagrams.} (a) Ising model; the dashed line is the $L^+$ crossover line. (b) Droplet picture for spin glasses (SG). (c) RSB picture for spin glasses; the solid line is the dAT line.
  }
\label{fig:PD}
\end{figure*}

The fundamental and unresolved challenge is to determine whether the dAT transition exists in 3D spin glasses. In this work, we provide direct numerical evidence for its existence in the 3D Edwards-Anderson (EA) spin-glass model through the analysis of a newly defined spin-glass susceptibility, guided by the following theoretical pictures.

{\it (i) Ising picture.}
In the standard Ising model without quenched disorder, the ground states are ferromagnetic (FM) rather than spin-glass. Applying a field \(h\) extends the zero-field critical point at \((T = T_{\rm c}, h = 0)\) into two {\it supercritical crossover lines}, denoted as \(L^\pm\) lines~\cite{li2024thermodynamic} (see Fig.~\ref{fig:PD}~a). These lines are defined by the loci of the maxima in the magnetic susceptibility \(\chi \equiv \partial m/\partial h\) for a constant $h>0$. Due to the \(Z_2\) symmetry, the two lines are symmetric about the \(T\)-axis (below we focus on the $L^+$ line for $h>0$). They obey a universal scaling law, \(h \propto (T - T_{\rm c})^\Delta\), where \(\Delta = \beta + \gamma\) is the gap exponent. This scaling has been established both theoretically and numerically in Refs.~\cite{li2024thermodynamic, li2025supercritical, lv2025quantum, hgwk-pzw2}. It can also be understood intuitively from the well-known scaling form \(\chi \propto |T - T_{\rm c}|^{-\gamma} \mathcal{X}\bigl(h / |T - T_{\rm c}|^\Delta \bigr)\), near $T_{\rm c}$. While this scaling function alone does not determine the sign of \(h/(T - T_{\rm c})\) along the \(L^+\) line, simulations of both 2D and 3D Ising models show that \(h/(T - T_{\rm c}) > 0\), meaning that the line curves to the right of the $T$-axis ({\it right-bending})~\cite{li2024thermodynamic}.

{\it (ii) Droplet picture}. The droplet theory~\cite{mcmillan1984scaling, fisher1987absence, fisher1988equilibrium, 10.1007/BFb0057515} assumes two pure spin-glass states in zero  field, which are related by a global spin flip. Below \( T_{\rm c} \), the system freezes into one of the two states. According to the droplet picture, no dAT line exists in any finite dimension. However, the scaling form of the spin-glass susceptibility,
$\chi_{q} \propto |T - T_{\rm c}|^{-\gamma} \, \mathcal{X}\bigl( h^{2} / |T - T_{\rm c}|^{\Delta} \bigr)$, implies the possibility of a crossover line (as in the Ising case), via the field-temperature scaling relation, $h^2 \propto |T - T_{\rm c}|^\Delta $. Here, \( h^{2} \) couples linearly to the spin-glass order parameter \( q \), and the critical exponents need not coincide with those of the Ising universality class. Again, the scaling relation alone does not determine whether the crossover line bends to the left or to the right in the \( h\)--\(T \) plane. Our numerical results for the 2D spin glass are consistent with the droplet picture and indicate a {\it right-bending} crossover line (see Fig.~\ref{fig:PD}~b; recall that $T_{\rm c} \to 0$ in the thermodynamic limit in 2D). Consequently, the droplet phase diagram remarkably resembles that of the Ising model.

{\it (iii) Replica symmetry breaking (RSB) picture.}
According to the RSB picture, the dAT line persists in finite dimensions~\cite{parisi1979infinite, parisi1980magnetic, mezard1987spin}. This line marks a true phase transition where the spin-glass susceptibility diverges. Below the dAT line, replica symmetry is broken among an infinite number of pure spin-glass states. Crucially, the dAT line must be {\it left-bending} in the $h$--$T$ plane, as it forms the boundary of the spin-glass phase that exists only below $T_{\rm c}$  (see Fig.~\ref{fig:PD}~c). Our numerical results show that the mean-field and 3D spin glasses are consistent with this picture.

Our strategy proceeds as follows. We simulate spin-glass models and compute the newly introduced spin-glass susceptibility, $\chi_W$, which exhibits a peak as a function of temperature $T$ at a fixed  $h$. The loci of these maxima, $h(T_{\rm max})$, trace out a line in the $h$-$T$ plane. If this line bends to the right, we can conclude that no dAT line exists in the system; this is what we observe in 2D. If, instead, the line bends to the left, it may signal dAT transitions, but it could be also crossovers (i.e., the spin-glass susceptibility remains finite in the thermodynamic limit).
Note that this criterion  is a qualitative  one, insensitive to numerical values.
To distinguish between a transition and a crossover, we further perform a finite-size analysis and check whether the peak value $\chi^{\rm max}_W$ of $\chi_W(T)$ diverges in the thermodynamic limit, for a fixed $h$. For the 3D EA model, we find that $\chi^{\rm max}_W$ increases with
the total number of spins $N$ as a power-law, indicating a true dAT transition in the thermodynamic limit. In contrast, for large 2D systems, $\chi^{\rm max}_W$ approaches a constant, consistent with a crossover. As a consistency check, we also verify our approach on the well-understood 3D Ising and mean-field spin-glass models.

\begin{figure*}[!htbp]
  \centering
  \includegraphics[width=\linewidth]{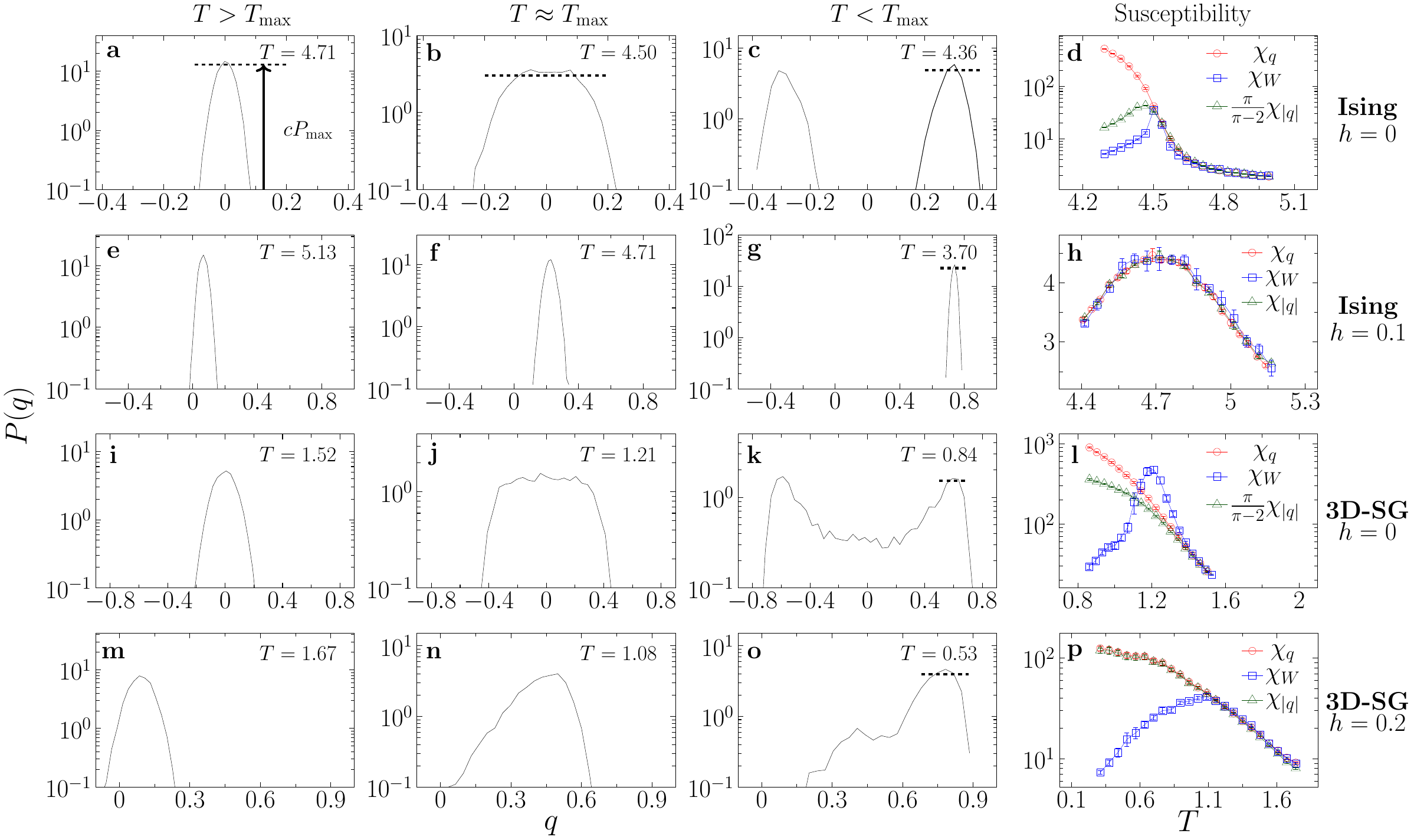}
\caption{{\bf Distribution of the overlap order parameter $P(q)$ and various susceptibilities.}
Data are obtained from simulations for the 3D Ising and spin-glass (SG) models ($L=16$).
Error bars represent the standard error of the mean, in all figures.
}
\label{fig:method}
\end{figure*}

\section{Models}
Simulations are implemented for models described by an Ising spin Hamiltonian ($\sigma_i = \pm 1$):
\begin{equation}
H = -\sum_{\langle ij \rangle} J_{ij}\sigma_i\sigma_j - h\sum_i \sigma_i.
\label{EA_hamiltonian_def}
\end{equation}
For the EA model~\cite{edwards1975theory}, the interactions $\langle ij \rangle$ are between nearest neighbors on a $d$-dimensional lattice (with $d=2$ and $3$ considered). For the mean-field version, the sum over $\langle ij \rangle$ includes all pairs of vertices on a Bethe lattice with a connectivity of $4$~\cite{parisi2012numerical, mezard2001bethe}. We employ two types of distributions for the couplings $J_{ij}$: a Gaussian distribution (mean $0$, variance $1$) for the $d=3$ case, and a binomial distribution with values $\pm 1$ occurring with equal probability for the $d=2$ and mean-field cases.

\section{A new method to compute the spin-glass susceptibility}
We perform Monte-Carlo simulations, utilizing the
parallel tempering technique~\cite{hukushima1996exchange} (see Appendix Sec.~\ref{sec:App_MC} and~\ref{sec:App_Equilibrium} for details).
The central observable is the overlap order parameter, $q \equiv \frac{1}{N}\sum_i \sigma^{A}_i\sigma^{B}_i$,
where $A$ and $B$ are two independent replicas of the system with identical bonds (quenched disorder). The distribution is  $P(q) \equiv [ \left \langle \delta \left (q - \frac{1}{N}\sum_i \sigma^{A}_i\sigma^{B}_i \right ) \right \rangle ]_{\rm av}$, and the standard spin-glass susceptibility is,   $\chi_{q} \equiv  N \left [ \left \langle q^2 \right \rangle - \langle q \rangle ^2   \right]_{\rm av}$,
where $\langle \dots \rangle$ denotes the thermal average, and $[ \dots ]_{\rm av}$ the average over disorder.

To motivate our consideration, it is useful to discuss two paradigmatic examples, the 3D Ising model and 3D spin glasses.

{\it (i) Ising model in zero field.}
The conventional order parameter for the Ising model is the magnetization,  $m \equiv \frac{1}{N}\sum_i \sigma_i$. However, to maintain consistency with spin-glass analyses, we will instead study the overlap order parameter, $q$, which, for the Ising model, is related to the magnetization via $q^2 \sim m^4$. The corresponding spin-glass susceptibility, $\chi_q = \partial q/\partial h$, is then proportional to $m \chi$ and therefore diverges at the critical temperature $T_{\rm c}$.

The distribution $P(q)$ has well-known behavior (Fig.~\ref{fig:method}~a–c): it is a Gaussian-like function above $T_{\rm c}$, and splits into two Gaussian-like peaks symmetric about $q=0$ below $T_{\rm c}$. If one computes $\chi_q$ from the variance of the full $P(q)$ as $\chi_q \equiv N(\langle q^2 \rangle - \langle q \rangle^2)$, then $\chi_q$ increases monotonically with decreasing $T$ (see Fig.~\ref{fig:method}~d). To capture the divergence at $T_{\rm c}$, a standard approach is to compute $\chi_{|q|} \equiv N(\langle |q|^2 \rangle - \langle |q| \rangle^2)$, which makes use of the time reversal symmetry. This shows that below $T_{\rm c}$, the correct susceptibility corresponds to the variance of a single peak in $P(q)$ - reflecting spontaneous symmetry breaking - rather than the variance of the full distribution. Above $T_{\rm c}$, $\chi_{q}$ and $\chi_{|q|}$ are proportional, with $\chi_{q} = \frac{\pi}{\pi-2} \chi_{|q|}$, a relation easily derived for a Gaussian distribution.

{\it (ii) Ising model in a field.} The application of a non-zero field ($h>0$) breaks time-reversal symmetry and suppresses the negative peak in $P(q)$ below $T_{\rm c}$. Since the distribution becomes very narrow, the probability of observing a negative $q$ is negligible in the temperature regime of interest near $T_{\rm c}$, making $\chi_{q}$ and $\chi_{|q|}$ effectively indistinguishable (see Fig.~\ref{fig:method}~e–h). The peak in $\chi_{|q|}(T)$ signals a supercritical crossover and has been used to locate the $L^\pm$ lines via the extracted $h(T_{\rm max})$ data. This approach yields the correct scaling $h \propto (T_{\rm max}-T_{\rm c})^\Delta$ in simulated Ising systems~\cite{li2021determining, lv2025quantum, hgwk-pzw2}.

As established, $\chi_{|q|}$ is sufficient to characterize both the phase transition at zero field and the supercritical crossovers under a finite field in the Ising model. However, this standard method fails for spin glasses due to the emergence of a multitude of metastable states, as we show below. The complex, multi-state structure of the spin-glass phase renders the trick of using the absolute value of the order parameter ineffective for isolating the relevant fluctuations, necessitating a novel approach to compute the susceptibility.

{\it (iii) Spin glass in zero field.} According to mean-field RSB theory~\cite{mezard1987spin}, below the critical temperature \(T_c\) and in the thermodynamic limit, the overlap distribution takes the form \(P(q) = c_1 \delta(q - q_{\rm EA}) + c_1 \delta(q + q_{\rm EA}) + \tilde{P}(q)\), where the Edwards-Anderson (EA) parameter \(q_{\rm EA}\) defines the maximum overlap, and \(\tilde{P}(q) > 0\) is a smooth function in the interval \(-q_{\rm EA} < q < q_{\rm EA}\). In the droplet picture, the continuous part \(\tilde{P}(q)\) vanishes in the thermodynamic limit, leaving only two $\delta$-peaks - resembling the \(P(q)\) of the Ising model. For finite-dimensional spin glasses, however, simulations show that \(\tilde{P}(q)\) remains significant in systems of finite size and does not disappear as the system size increases~\cite{katzgraber2001monte, yucesoy2012evidence} (see  Fig.~\ref{fig:method}~i-l and Fig.~\ref{fig:method_appendix}, in the Appendix). Because of the contribution from \(\tilde{P}(q)\), the susceptibility \(\chi_{|q|}(T)\) varies monotonically and shows no peak (Fig.~\ref{fig:method}~l)~\cite{bhatt1988numerical}, in contrast to the pronounced peak observed in the Ising model (Fig.~\ref{fig:method}~d).

{\it (iv) Spin glass in a field.} When \(h>0\), the RSB theory predicts~\cite{mezard1987spin} in the spin-glass phase below the dAT line and in the thermodynamic limit,
\(
P(q) = c_0 \delta(q - q_0) + c_1 \delta(q - q_{\rm EA}) + \tilde{P}(q),
\)
where \(\tilde{P}(q) > 0\). In contrast, the droplet picture predicts a single $\delta$-function, \(P(q) \sim \delta(q - q_{\rm EA})\), analogous to the Ising model.
In simulated finite-size systems, the tail \(\tilde{P}(q)\) is typically very broad and yields non-negligible probability at negative \(q\) values, making \(\chi_{|q|}\) problematic (see  Fig.~\ref{fig:method}~m-p and Fig.~\ref{fig:method_appendix}, in the Appendix). Moreover, \(\chi_{|q|}(T)\) also fails to exhibit a peak. In short, $\chi_q$ and  \(\chi_{|q|}\) are  unsuitable for spin glasses, both with and without a field.

To overcome the aforementioned difficulties, we require a new method to estimate the spin-glass susceptibility.
The lesson from the Ising model is that, within the ordered phase, the susceptibility should be derived from the variance of a single peak, rather than from the full distribution. The critical behavior at \(T_{\rm c}\) is characterized by the vanishing curvature (second derivative) of this peak.  In the spin-glass phase and for sufficiently large systems, the \(\delta\)-function-like peak in \(P(q)\) can be separated from the continuous part  \(\tilde{P}(q)\). This motivates defining the following quantity, \(\chi_W\), which captures the local curvature of the peak while minimizing the influence from the broad tail.

We define the width \( w = |q_{2} - q_{1}| \) of a peak, where \( P(q_{1}) = P(q_{2}) = c P_{\text{max}} \) with \( 0 < c < 1 \), and \( P_{\text{max}} \) is the maximum value of \( P(q) \) (see Fig.~\ref{fig:method}~a). The susceptibility is then defined as \( \chi_{W} = N w^2 \).
The fraction \( c \) is chosen to minimize the mean squared difference between \( \chi_W \) and the standard susceptibility \( \chi_q \) over \( M \) data points at a series of temperatures \( T_i \):
\(
c = \arg\min_{c'} \left\{ \frac{1}{M} \sum_{i=1}^{M} \left( \chi_W(c', T_i) - \chi_q(T_i) \right)^2 \right\}.
\)
The optimization uses temperatures \( T > k T_{\text{max}} \), where \( k \in (1, 1.2) \) has a slight model and parameter dependence (see Table~\ref{table:coefficients}).
As a general principle, we require \( \chi_W \) and \( \chi_q \) to match in the high-temperature phase (see Appendix Sec.~\ref{sec:Optimization_parameters} for further details on the determination of $k$ and $c$).
 Typically, \( c \) is around 0.88, indicating that we monitor the broadening of the peak very close to its maximum (see Fig.~\ref{fig:method}(k,o)).
Importantly, this method minimizes the influence of the broad tail at low temperatures, resulting in a peaked \( \chi_W(T) \) for all cases studied, regardless of the model or the presence of a field \( h \). Examples for the Ising and 3D spin-glass models are shown in Fig.~\ref{fig:method}, with additional examples for 2D and mean-field spin glasses provided in the Appendix.

\begin{figure}[!htbp]
  \centering
  \includegraphics[width=1\linewidth]{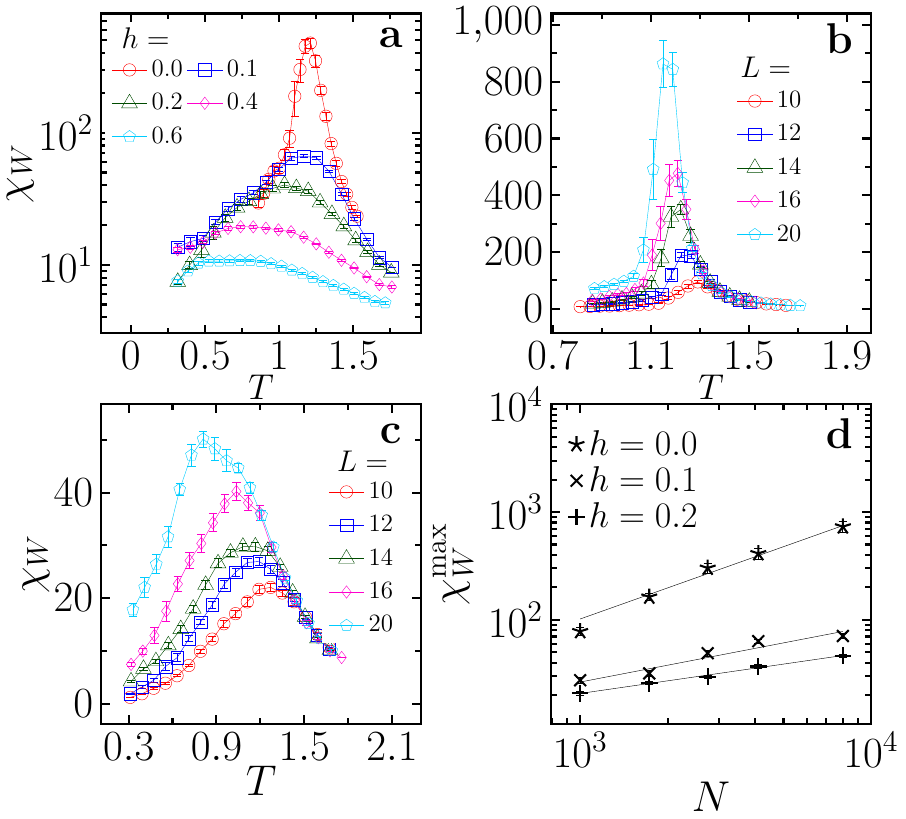}
  \caption{{\bf Results for the 3D spin glasses.} (a-c) Data of susceptibility $\chi_W(T)$: (a)  $L= 16$, (b) $h=0$  and (c) $h=0.2$. (d) Peak value $\chi_W^{\rm max}$ as a function of $N=L^3$. The lines represent power-law fitting, $\chi_W^{\rm max} \sim N^{\alpha}$.
  }
  \label{fig:finite_size3D}
\end{figure}

\begin{figure}[hbt!]
  \centering
  \includegraphics[width=1\linewidth]{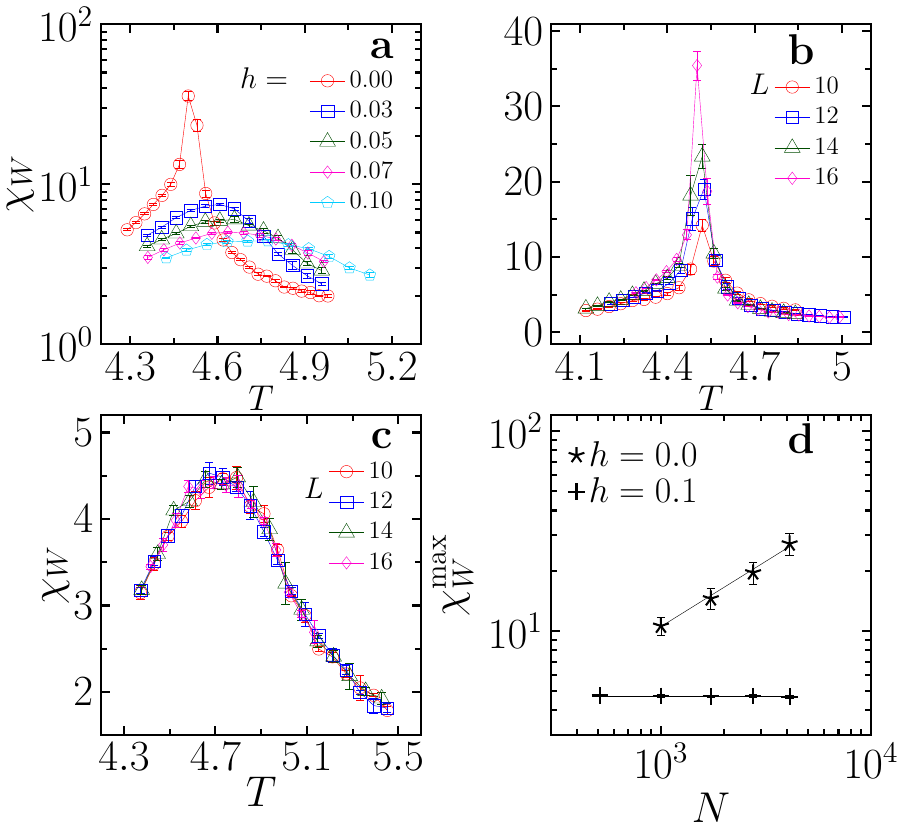}
  \caption{{\bf Results for the 3D Ising model.} (a-c) Data of susceptibility $\chi_W(T)$: (a)  $L= 16$, (b) $h=0$, and (c) $h=0.1$. (d) Peak value $\chi_W^{\rm max}$ as a function of $N=L^3$. The lines represent $\chi_W^{\rm max} \sim N^{\alpha}$, with  $\alpha = (2-\eta)/d \approx 0.65$ for $h=0$ and $\alpha = 0$ for $h=0.2$.}
\label{fig:finite_sizeIsing}
\end{figure}

\section{Results}

\subsection{Dependence of the locus $T_{\rm max}$ of the susceptibility maximum  on the field $h$}
\begin{figure}[hbt!]
  \centering
  \includegraphics[width=1\linewidth]{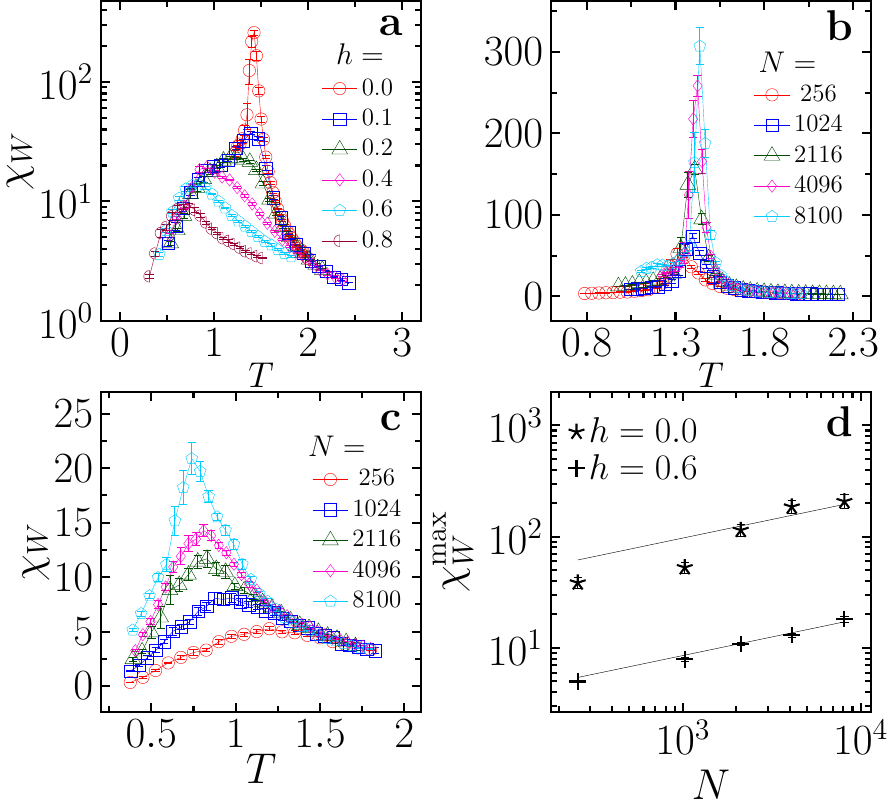}
  \caption{{\bf Results for the mean-field spin-glass model.} (a-c) Data of susceptibility $\chi_W(T)$: (a) $N=4096$, (b) $h=0$, and (c) $h=0.6$. (d) Peak value $\chi_W^{\rm max}$ as a function of $N$. The lines represent $\chi_W^{\rm max} \sim N^{1/3}$.}
\label{fig:finite_mf}
\end{figure}

\begin{figure}[hbt!]
  \centering
  \includegraphics[width=1\linewidth]{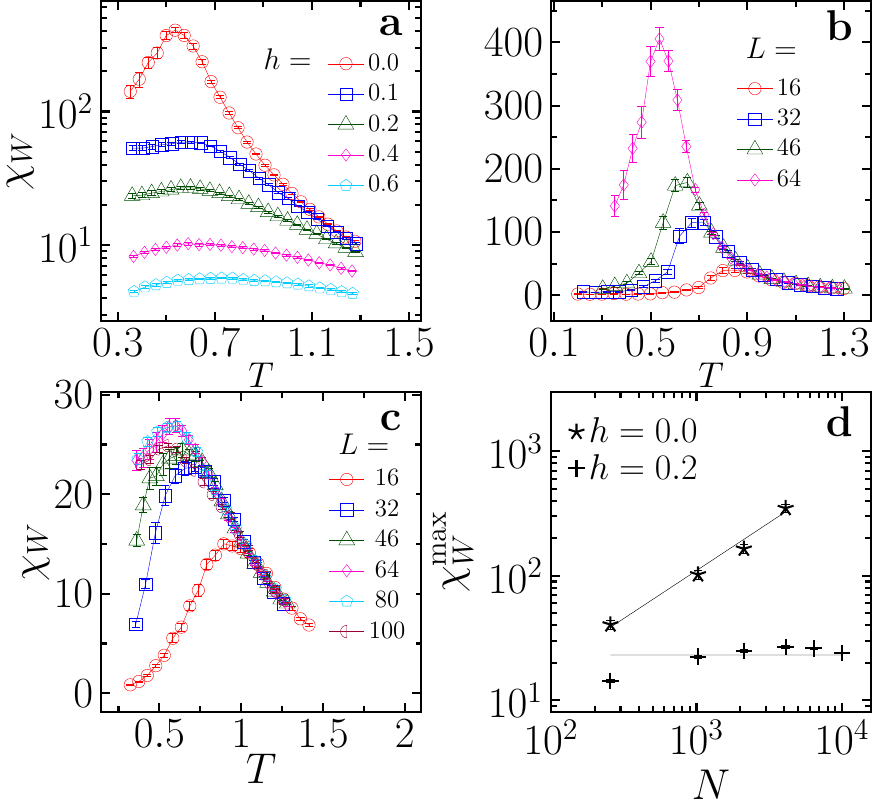}
  \caption{{\bf Results for the 2D spin glasses.} (a-c) Data of susceptibility $\chi_W(T)$: (a) $L=64$, (b) $h=0$, and (c) $h=0.2$. (d) Peak value $\chi_W^{\rm max}$ as a function of $N=L^2$. The lines represent, $\chi_W^{\rm max} \sim N^{\alpha}$, with  $\alpha \approx 0.77$ for $h=0$ and $\alpha = 0$ for $h=0.2$.
  }
\label{fig:finite_size2D}
\end{figure}

\begin{figure}[!htbp]
  \centering
  \includegraphics[width=1\linewidth]{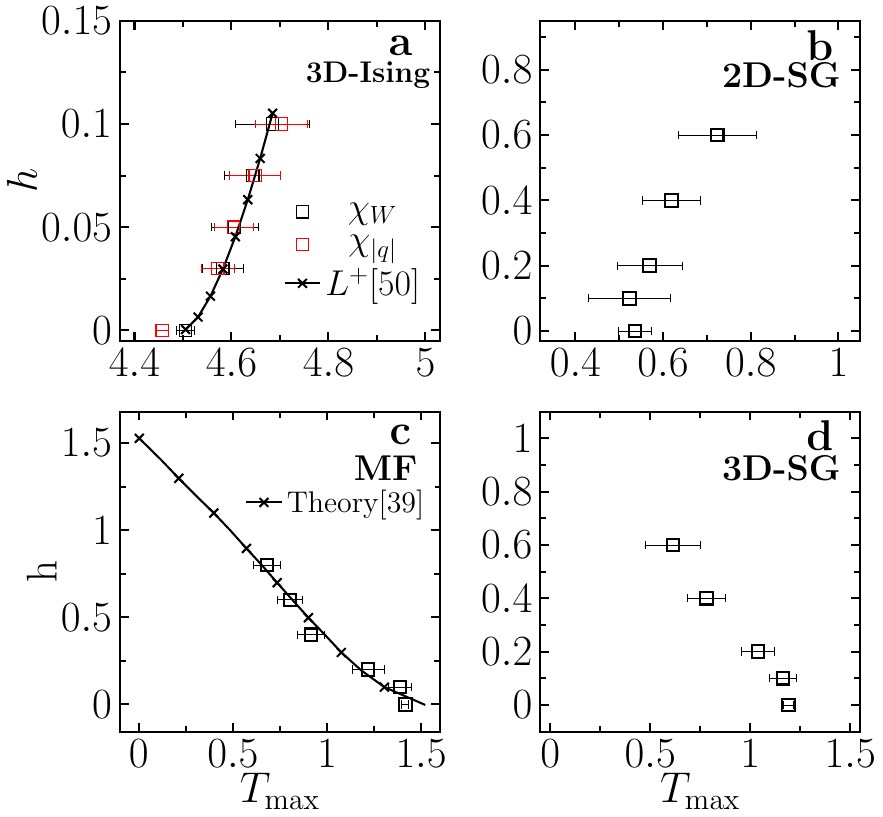}
  \caption{{\bf Results of $h(T_{\rm max})$ for four models, determined by the loci  of maximum $\chi_W(T)$.}
  In (a), the loci of maximum $\chi_{|q|}(T)$  are added for comparison.}
\label{fig:dAT}
\end{figure}

The curves of $\chi_W(T)$ at several $h$ are shown in Fig.~\ref{fig:finite_size3D}~a for the 3D spin-glass model.
Similarly, the Figs.~\ref{fig:finite_sizeIsing},~\ref{fig:finite_mf},~\ref{fig:finite_size2D} display results for 3D Ising, mean-field, and 2D spin-glass models, respectively. A bootstrap procedure is used to extract the peak position $T_{\rm max}$ and the peak value $\chi_W^{\rm max}$ (see Appendix Sec.~\ref{sec:App_Bootstrap}). The dependence of $T_{\rm max}$ on $h$ is plotted in Fig.~\ref{fig:dAT} for the four models studied.

Two distinct groups are clearly observed: for the 3D Ising and 2D spin-glass models, the $h(T_{\rm max})$ curves bend to the right, whereas for the mean-field and 3D spin-glass models, they bend to the left. The $h(T_{\rm max})$ curve for the 3D Ising model is consistent with the previously estimated $L^+$ supercritical crossover line~\cite{li2024thermodynamic}.  Since the curve for the 2D spin-glass model bends to the right, we conclude that no dAT line exists in two dimensions, in agreement with established knowledge. For the mean-field spin-glass model, the $h(T_{\rm max})$ data align closely with the theoretical dAT line~\cite{parisi2012numerical} (Bethe lattice with a connectivity of $4$). The similar left-bending behavior of $h(T_{\rm max})$ in the 3D spin-glass model suggests the possible presence of a dAT line in three dimensions.

\subsection{Finite size analysis of the maximum susceptibility $\chi_W^{\rm max}$.}

The dAT line marks the stability limit of the replica--symmetric solution~\cite{de1978stability}, and separates the high--temperature paramagnetic phase from the low--temperature spin--glass phase. In the thermodynamic limit, the usual spin--glass susceptibility $\chi_q$ diverges in the spin--glass phase and remains finite in the paramagnetic phase. In contrast, the new susceptibility $\chi_W$, which probes the curvature of the peak in $P(q)$, behaves differently: we expect it to diverge only at the dAT line, while staying finite in both phases. Hence, a finite--size scaling analysis of $\chi_W^{\mathrm{max}}$ provides a useful diagnostic---if the scaling suggests divergence of $\chi_W^{\mathrm{max}}$  in the thermodynamic limit, this signals the existence of a dAT transition.

To examine finite-size effects, we plot  $\chi_W(T)$ for various system sizes ($N = L^d$) in the 3D spin-glass model, at  $h=0, 0.1$ and $0.2$ (Fig.~\ref{fig:finite_size3D}~b,c and Fig.~\ref{fig:finite_size3D_h010}, in the Appendix). The peak values $\chi_W^{\rm max}$ exhibit power-law scaling, $\chi_W^{\rm max} \sim N^{(2-\eta)/d}$ (Fig.~\ref{fig:finite_size3D}~d). Fitting this relation yields $\eta = -0.9(3), 0.4(1), 0.83(1)$ for $h=0, 0.1, 0.2$ respectively (see Table~\ref{table:coefficients}).
For the zero-field case ($h=0$), our estimate of $\eta = -0.9(3)$ is consistent with previously reported values that are in the range $-0.6 < \eta < -0.2$~\cite{katzgraber2006universality, ballesteros2000critical, baity2013critical}. 

For the standard 3D Ising model (see  Fig.~\ref{fig:finite_sizeIsing}) at $h=0$, $\chi_W^{\rm max}$ follows the expected finite-size scaling with the Ising universality exponent $\eta \approx 0.036$, whereas for $h=0.1$, $\chi_W^{\rm max}$ becomes approximately constant, signaling a crossover. In the mean-field spin-glass model, our data conform to the theoretical scaling $\chi_W^{\rm max} \sim N^{1/3}$ at both $h=0$ and $h=0.6$ (see  Fig.~\ref{fig:finite_mf})~\cite{parisi1993critical, billoire2011finite, aspelmeier2016finite}.
For the 2D spin-glass model (see  Fig.~\ref{fig:finite_size2D}) at $h=0$, we find $\eta = 0.46(2)$, close to the range of $0.2 < \eta < 0.4$ reported in prior studies~\cite{morgenstern1980magnetic, mcmillan1983monte, bhatt1988numerical, 10.1007/BFb0057515}; in contrast, at $h=0.2$, $\chi_W^{\rm max}$ saturates to a constant for $L \geq 32$, again indicating a crossover.

The chosen field strengths $h$ in Table~\ref{table:coefficients} match those previously reported in the literature, in the corresponding models. Since $T_{\rm max}$ decreases with the increasing $h$ for the mean-field and 3D spin-glass models, we restrict our simulations to moderate field values due to numerical constraints. Simulations at larger $h$ would require equilibration at lower temperatures in order to obtain the peak in $\chi_W(T)$, a regime that is currently inaccessible.

In summary, our data are in general agreement with established results for the Ising, 2D spin-glass and mean-field spin-glass models (see Table~\ref{table:coefficients} for a summary). This consistency validates the present approach and reinforces our main conclusion regarding the existence of dAT transitions in three dimensions.

\begin{table}[h]
\centering
\caption{Values of $c$, $k$, $\alpha$ and $\eta$.
For crossovers, $\alpha$ and $\eta$ are indicated by ``-''.
}
\begin{tabular}{c | c | c c c c}
\hline
\hline
\qquad \qquad & $h$ & $c$ & $k$ & $\alpha$ & $\eta$\\
\hline
\multirow{2}{*}{Ising}  & 0  & 0.87 & 1.03  &  0.65 & 0.036 \\
  & 0.1  & 0.88 & 1 & - & - \\
  \hline
\multirow{2}{*}{MF SG}  & 0  & 0.88 & 1.13  & 1/3 & 0 \\
 & 0.6  & 0.87 & 1.12 & 1/3 & 0 \\
\hline
\multirow{2}{*}{2D SG}  & 0  & 0.88 & 1.19  &  $0.77\pm 0.08$  & $0.46\pm0.2$\\
 & 0.2  & 0.88 & 1.05 & - & - \\
\hline
\multirow{3}{*}{3D SG}  & 0    & 0.88 & 1.17 & $0.96\pm 0.09$   &  $-0.9 \pm 0.3$  \\
                        & 0.1  & 0.88 & 1.17 & $0.52\pm 0.05$    &  $0.4 \pm 0.1$  \\
                        & 0.2  & 0.88 & 1.07 & $0.390\pm 0.004$  &  $0.83 \pm 0.01$  \\

\hline
\hline
\end{tabular}
\label{table:coefficients}
\end{table}

\section{Discussion}
Our conclusion is constrained by the range of simulated system sizes for 3D spin-glass models; $L=20$ represents the maximum feasible with commodity hardware. To perform a more extensive finite-size scaling analysis, such as for the data in Fig.~\ref{fig:finite_size3D}~d, simulations on specialized Janus computer~\cite{baity2014janus} - capable of generating equilibrium configurations with $L > 30$~\cite{baity2014three} - would be beneficial. 
Alternative approaches to overcome this computational bottleneck include implementing parallel tempering on GPU clusters and employing the recently developed cluster-move algorithm~\cite{chilin2026cluster}. Generally, reliable conclusions can be drawn only when the scaling regime in Fig.~\ref{fig:finite_size3D}~d extends over at least two decades — a requirement that sets a clear target for future studies.

Another important future direction is the systematic estimation of critical exponents across dimensions, to be compared with the field theoretical results~\cite{charbonneau2017nontrivial}.
Alternatively, the current method can be applied to 1D spin chains with long-range interactions~\cite{angelini2026long}, which are often computationally more tractable. These 1D long-range models serve as a standard proxy for higher-dimensional short-range spin glasses, as the  decay exponent $\rho$ in the long-range interaction maps directly onto the dimensionality $d$.
Finally, it is straightforward to apply the current method to the Gardner transition in structural glasses~\cite{charbonneau2014fractal, berthier2016growing, urbani2023gardner}, which belongs to the same universality class of the dAT transition.

\section*{Acknowledgments}

We warmly thank Patrick Charbonneau, Dmytro Khomenko, Xinyang Li, Hajime Yoshino, Deng Pan and Adauto de Souza for inspiring discussions. This work was supported by National Key R\&D Program of China (Grant No. 2025YFF0512000), the Space Application System Of China Manned Space Program, Wenzhou Institute (Grant No. WIUCASICTP2022), and the National Natural Science Foundation of China (Grant No. 12447101). We acknowledge the use of the High Performance Cluster at Institute of Theoretical Physics - Chinese Academy of Sciences, and the computer clusters at Hefei advanced computing center.


\appendix

\section{Monte-Carlo (MC) simulations with parallel tempering acceleration}\label{sec:App_MC}
The parallel tempering technique alternates between multiple canonical processes at different temperatures, swapping configurations (replicas) based on the total system energy cost~\cite{hukushima1996exchange}.
Sample exchanges occur with a considerable probability, ensuring that the entire temperature range is visited~\cite{hukushima1996exchange}.
To circumvent free energy barriers, replicas must cover rapid-equilibrium and slow-relaxation regimes.
Thus, our temperature distributions span both sub- and supercritical values\cite{katzgraber2006feedback}.

Temperature-space bottlenecks are mitigated by centering Gaussian distributions of inverse temperatures on regions exhibiting low replica-exchange acceptance rates.
We then adjust these distributions to obtain uniform exchange probabilities among all replica pairs~\cite{rozada2019effects}. The exchanges only involve adjacent replicas.
Since energy fluctuations affect exchange acceptance rates, the required number of replicas scales with system size.
In the present study, this number ranges from 8 (for small system sizes) to 48 (for the largest system sizes). The unit of time is one MC step, defined as $N$ spin-flip attempts.\\

\begin{figure}[hbt!]
  \centering
  \includegraphics[width=\linewidth]{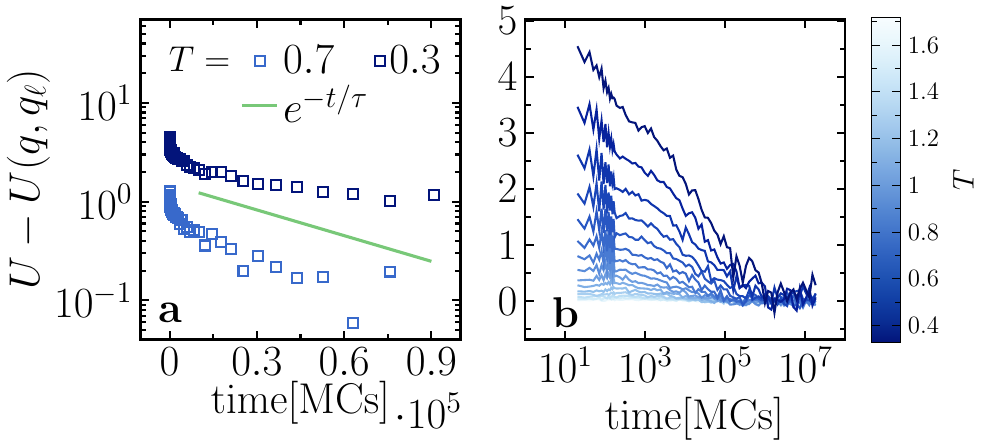}
    \caption{{\bf Equilibrium test.}
    (a) Relaxation dynamics of the difference between the directly computed energy $U$ and  $U(q, q_\ell)$, for 3D spin glasses ($L=14$, $h=0.2$).
    Data are compared to an exponential function with $\tau = 2 \times 10^5$ (line). (b) Temperature-dependent relaxation (indicated by color). Results are averaged over 4 independent $J_{ij}$ realizations, each with 4 replicas.
    }
  \label{fig:rlx}
\end{figure}

\section{Equilibrium criterion}\label{sec:App_Equilibrium}

The relaxation time $\tau$ is estimated from the standard equilibrium test~\cite{young2004absence,katzgraber2006universality,hukushima1998exchange},
which consists of comparing the directly computed energy $U$ with the energy expressed via the average site overlap ($q$) and link overlap ($q_\ell$):
\begin{equation}
  U(q, q_\ell) = \frac{z}{2}\,\frac{q_\ell-1}{T} + \frac{q-1}{T}h^2,
  \label{UqABql_def}
\end{equation}
where $z$ is the number of neighbors and $q_\ell = \frac{2}{zN}\sum_{\langle ij \rangle} \langle\sigma^{A}_i\sigma^{A}_j \sigma^{B}_i\sigma^{B}_j \rangle$ is the link overlap.

An exponential fit of $U - U(q, q_\ell)$ as a function of time provides an estimate for the relaxation time $\tau$ (see Fig.~\ref{fig:rlx}a). For instance, the relaxation time $\tau$ is $\sim 10^5$ MC steps for the 3D spin glass with $L=14$ and $h=0.2$. Simulations are performed for a duration of 10$\tau$ to 100 $\tau$, to ensure that $U - U(q, q_\ell)$ vanishes and the system reaches equilibrium (see Fig.~\ref{fig:rlx}b).

Using equilibrated samples, the spin overlap $q$ is averaged over time. The histogram $P(q)$ is then constructed from both replicas and realizations of disorder $\{ J_{ij} \}$.
We use 30 independent $\{ J_{ij} \}$ realizations, each with 20 replicas, to obtain the results in the main text.

\begin{figure}[hbt!]
  \centering
  \includegraphics[width=0.7\linewidth]{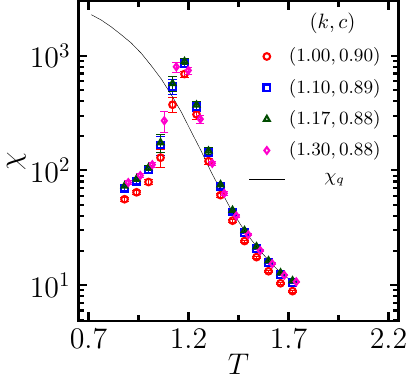}
  \caption{\textbf{Optimization of $(k,c)$.} Simulation data for the 3D spin-glass  model with $L=20$ at $h=0$. The solid line represents $\chi_q$, while the symbols show $\chi_W$ for selected cutoff factors $k = 1,\,1.10,\, 1.17,$ and $1.30$. The  parameter $c$ is obtained from best fitting for the given $k$.}
\label{fig:Optimization_parameters}
\end{figure}

\begin{figure*}[hbt!]
  \centering
  \includegraphics[width=0.95\linewidth]{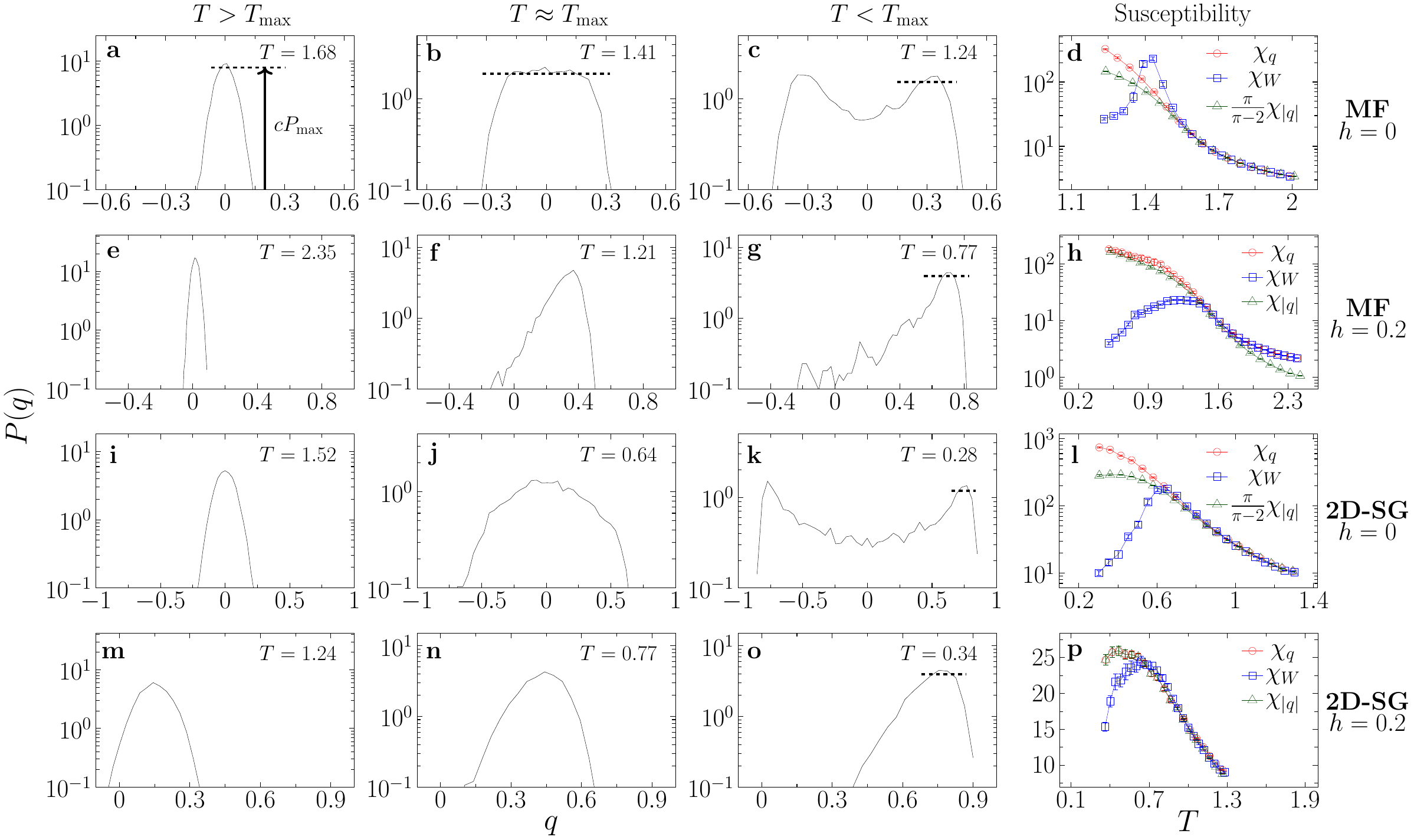}
\caption{{\bf Distribution of the overlap order parameter $P(q)$ and various susceptibilities.}
Data are obtained from simulations for the mean-field (MF) and 2D spin-glass (SG) models ($N=4096$ and $L=46$, respectively).
}
\label{fig:method_appendix}
\end{figure*}

\section{Optimization of the fractional height $c$ and cutoff factor $k$}\label{sec:Optimization_parameters}

The parameters $c$ and $k$ are determined using the following procedure. The cutoff factor $k$ is initially set to unity ($k=1$), and $c$ is then obtained by fitting $\chi_W$ to $\chi_q$ over the temperature range $T > k T_{\rm max}$. We then slightly increase $k$ and repeat the fitting to obtain an optimized $c$ for each given $k$. Figure~\ref{fig:Optimization_parameters} shows that the resulting $\chi_W$ depends only weakly on $k$ and converges rapidly as $k$ increases, with the $\chi_W$ data collapsing for $1.1 \leq k \leq 1.3$. As a general principle, we select the minimum $k$ such that further increases in $k$ do not appreciably change either $\chi_W$ or its deviation from $\chi_q$. 

\section{Bootstrap procedure to determine the peak of $\chi_W(T)$}\label{sec:App_Bootstrap}

A bootstrap procedure estimates $\chi_W^{\text{max}}$ and its correspondent $T_{\rm max}$ from the data of $\chi_W(T)$.
As 8–48 replicas are insufficient for a reliable peak determination, the parallel tempering procedure is repeated to accumulate hundreds of temperature points (ranging from $\sim 150$ to $\sim 550$).
Then, we combine interpolation with resampling, averaging the peak values over approximately $10^3$ bootstrap iterations.

\section{Distribution of overlap order parameter $P(q)$ for mean-field and 2D spin-glass models}

Figure~\ref{fig:method_appendix} shows the distribution of overlap order parameter $P(q)$ and corresponding susceptibilities for mean-field and 2D spin-glass models.

\section{Additional simulation results for 3D spin glasses}\label{sec:App_results}

\begin{figure}[hbt!]
  \centering
  \includegraphics[width=0.8\linewidth]{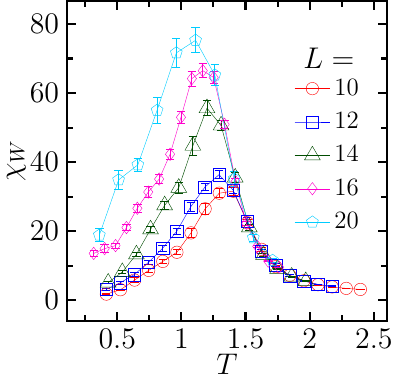}
  \caption{{\bf Susceptibility $\chi_W(T)$ at $h=0.1$ for the 3D spin-glass model.}}
  \label{fig:finite_size3D_h010}
\end{figure}

\begin{figure}[hbt!]
  \centering
  \includegraphics[width=0.8\linewidth]{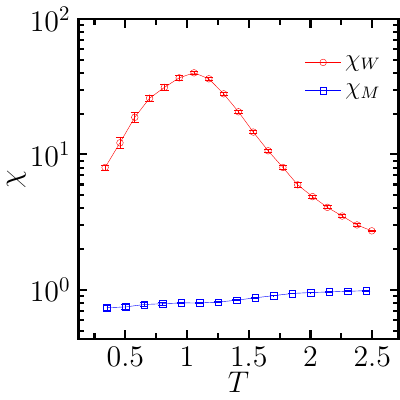}
  \caption{{\bf Data of susceptibilities $\chi_W(T)$ and $\chi_M(T)$ with $h=0.2$ and $L=16$, for the 3D spin-glass model.}}
  \label{fig:chiM}
\end{figure}

Simulation results for $\chi_W(T)$ in the 3D spin-glass model under $h=0.1$ are presented in Fig.~\ref{fig:finite_size3D_h010}.
We also plot $\chi_W(T)$ alongside $\chi_M(T)$ extracted from magnetization for $h=0.2$, in Fig.~\ref{fig:chiM}.
No peak is observed in $\chi_M(T)$, suggesting the absence of a mixed phase where ferromagnetic and spin-glass order coexist~\cite{de1978stability}. 

\clearpage


\bibliography{main_pre}

\end{document}